\documentclass[useAMS,usenatbib,usegraphicx]{mn2e}

\usepackage{times}

\newcommand{\figcolsize}{0.25}

\newcommand{\araa}{ARA\&A}   
\newcommand{\aj}{AJ}         \newcommand{\apj}{ApJ}
      \newcommand{\apjs}{ApJS}
\newcommand{\mnras}{MNRAS}   
     \newcommand{\pasp}{PASP}

\title[What Hubble really meant by late and early type]
{What Hubble really meant by late and early type:
  simply more or less complex in appearance}
\author[Ivan K. Baldry]{Ivan K. Baldry \\
Astrophysics Research Institute, Liverpool John Moores University, 
  Twelve Quays House, Egerton Wharf, Birkenhead CH41~1LD, UK}

\begin{document}

\date{Submitted 08/08/08 to Astronomy \& Geophysics;
        revised 2008 August 29.}

\pagerange{\pageref{firstpage}--\pageref{lastpage}} \pubyear{2008}

\maketitle

\label{firstpage}

\begin{abstract}
It is widely written and believed that Edwin Hubble introduced the terms
`early' and `late types' to suggest an evolutionary sequence for
galaxies. This is incorrect. Hubble took these terms from spectral
classification of stars to signify a sequence related to complexity of
appearance, albeit based on images rather than spectra. The temporal
connotations of the terms had been abandoned prior to his 1926 paper on
classification of galaxies.
\end{abstract}


\section{Introduction}

The terms `early' and `late type' in astrophysics have been applied to both
stars and galaxies.  Spectral classification of stars follows an early-to-late
sequence O-B-A-F-G-K-M with recent additions of L-T \citep{kirkpatrick99}.
This classification closely relates to a sequence in temperature from hot to
cool stars. Morphological classification of galaxies is based on a number of
factors, including ellipticity, the size of the nuclear region relative to the
spiral arms, and the smoothness of the image. 
A commonly used classification is the revised and extended Hubble
system \citep{deVaucouleurs59,Sandage61,Sandage75} that is based on Hubble's
(\citeyear{Hubble26}) original scheme for `extragalactic nebulae', and
Reynold's (\citeyear{Reynolds20}) earlier ideas.  It follows 
an early-to-late sequence, ellipticals-lenticulars-spirals-irregulars,
E-S0-Sa-Sb-Sc-Sd-Sm-Im (ignoring the barred/unbarred characteristic).
\citet{Sandage05} has reviewed the history of this development.

At first glance, there appears to be no relation between the early to late
type sequences of stars and galaxies other than the terminology. Before
the implications of $E=mc^2$ \citep{Einstein05} were realised, and to explain the 
Hertzsprung-Russell diagram, it was natural to suppose that stars cooled from
early to late spectral types because there was no established mechanism for
Myr stability of stellar atmospheres (cf.\ cooling of brown dwarfs,
\citealt{burrows97}). There is now a wide-spread but mistaken
belief that Hubble chose this
terminology because he thought that the morphological sequence was also a
temporal sequence.  For example, \citet{BM98book} wrote ``Hubble suggested
that galaxies evolved from the left-hand end of this sequence to the
right. This now discredited speculation lives on in the convention
... early-type ... late-type galaxies.''  While, \citet{CL02book} noted
``Although it is now not thought this evolutionary sequence is correct,
Hubble's nomenclature, in which ellipticals are `early' type and spirals and
irregulars `late', is still commonly used.''  Similar explanations can be
found in other textbooks (e.g.,
\citealt{Tayler93book,Shore03book,CO06book}). The main aim of this article is
to show that these explanations for the origin of the terminology are
incorrect, and to illuminate the correct explanation.

\section{Sequences in complexity of appearance}

The temporal meanings of `early' and `late' were questioned
for stellar spectra
by the early 1920's because of, for example, the discovery of red giants and
the suggestion of a nuclear timescale by
\citet{Eddington19}. \citet{Stratton24} quotes a 1922 International
Astronomical Union report ``The terms ... are very convenient. It is well,
however, to emphasize that they denote positions early or late in the spectral
sequence ... without any necessary connection whatever with an early or late
stage of physical evolution.''  Responding to a suggestion that the terms be
dropped by \citet{Hepburn24}, presciently \citeauthor{Stratton24} said ``it
may be doubtful whether words so strongly entrenched in the literature of the
subject can now be displaced ...'' In fact they have not been.

In Hubble's 1926 paper on the morphological sequence of galaxies the footnote
on page~326 is revealing: ``Early and late, in spite of their temporal
connotations, appear to be the most convenient adjectives for describing
relative positions in the sequence. ... They can be assumed to express a
progression from simple to complex forms.  An accepted precedent for this
usage is found in the series of stellar spectral types. There also the
progression is ... from the simple to the complex ... the temporal
connotations ... have been deliberately disregarded.'' Furthermore,
\citet{Hubble27} noted ``The nomenclature, it is emphasized refers to position
in the sequence, and temporal connotations are made at one's peril. The entire
classification is purely empirical and without prejudice to theories of
evolution ...''

\begin{figure*}
\centerline{
\includegraphics[width=\figcolsize\textwidth]{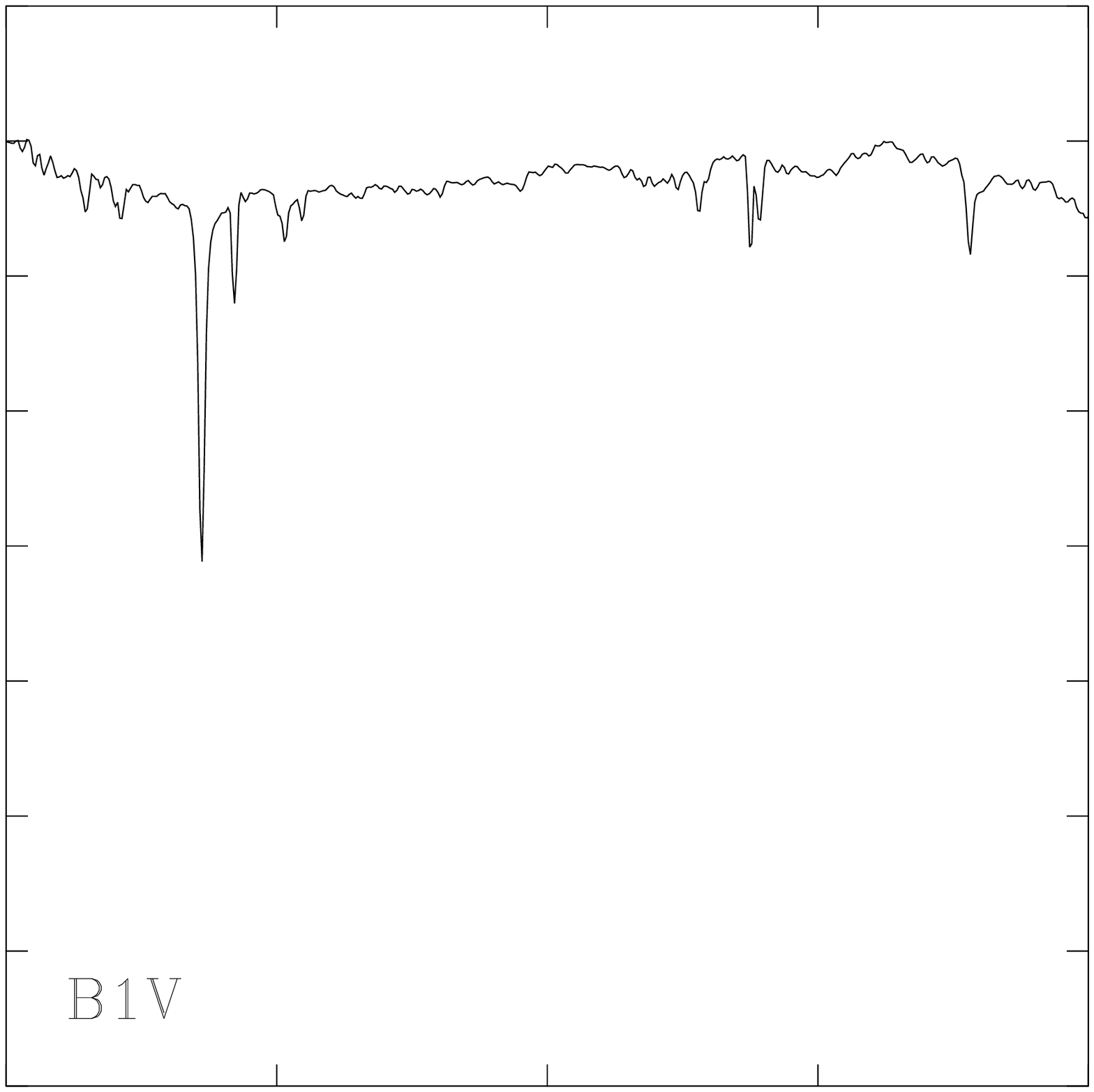}
\includegraphics[width=\figcolsize\textwidth]{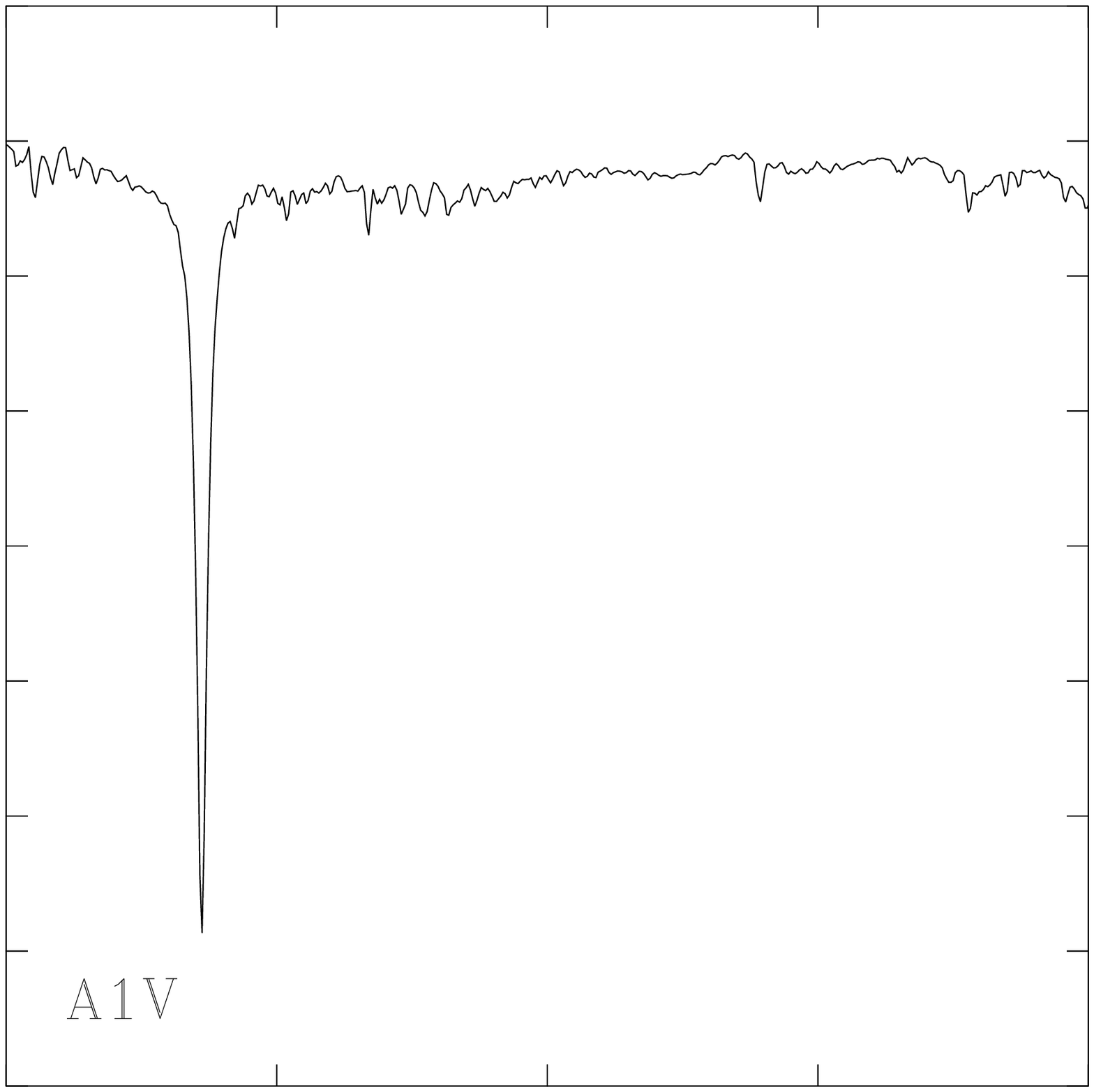}
\includegraphics[width=\figcolsize\textwidth]{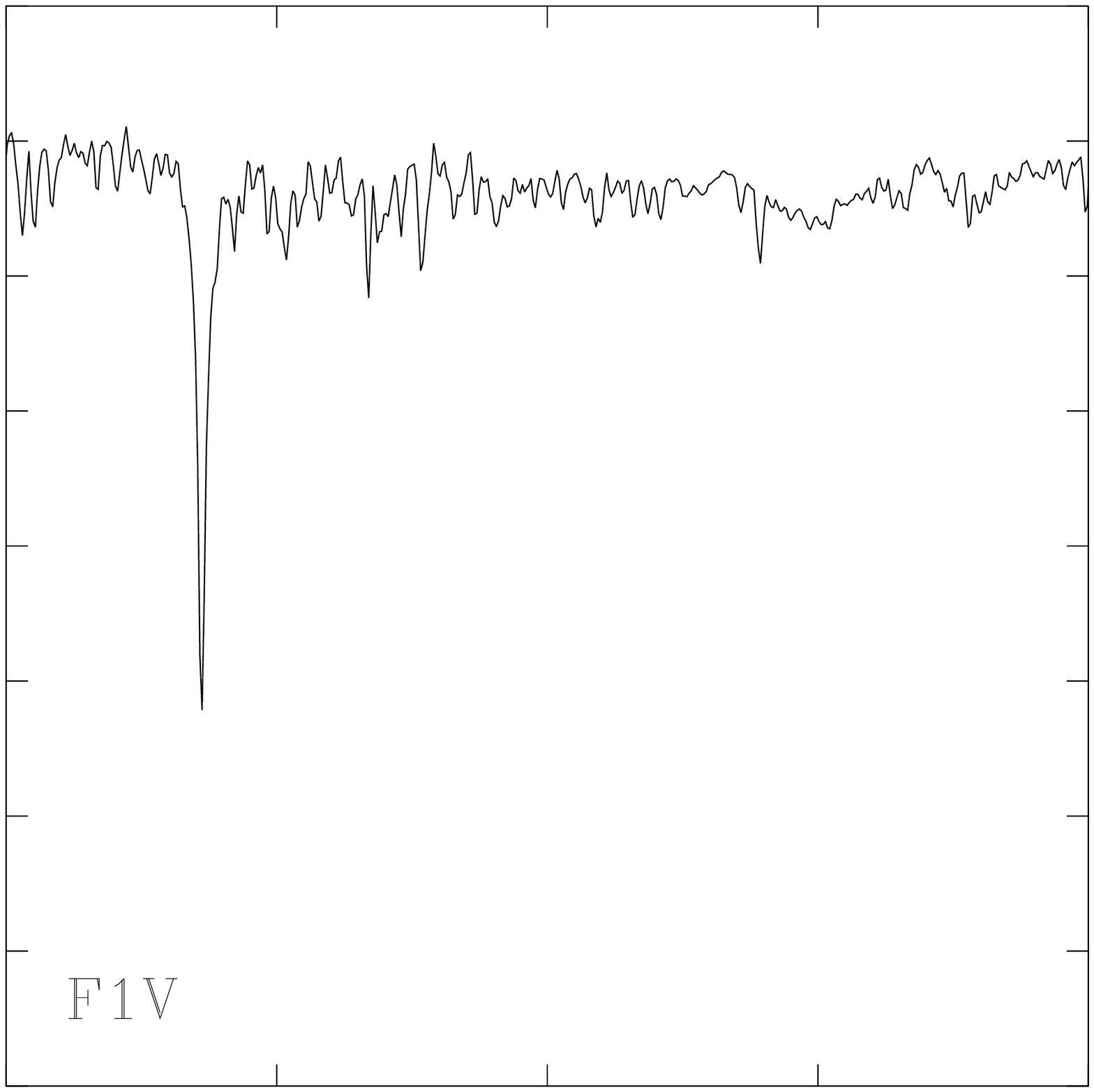}
\includegraphics[width=\figcolsize\textwidth]{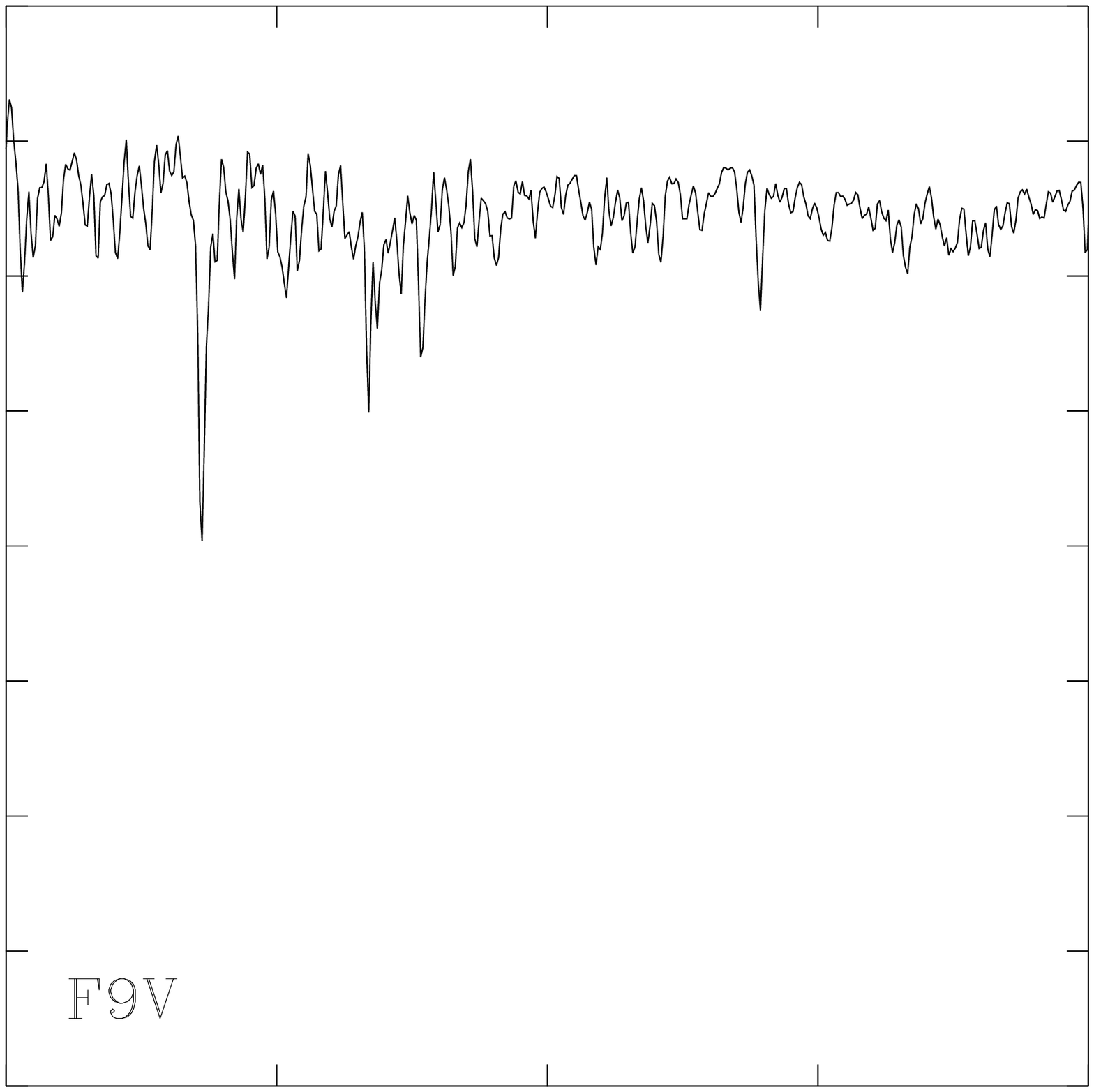}
}
\vspace{0.4cm}
\centerline{
\includegraphics[width=\figcolsize\textwidth]{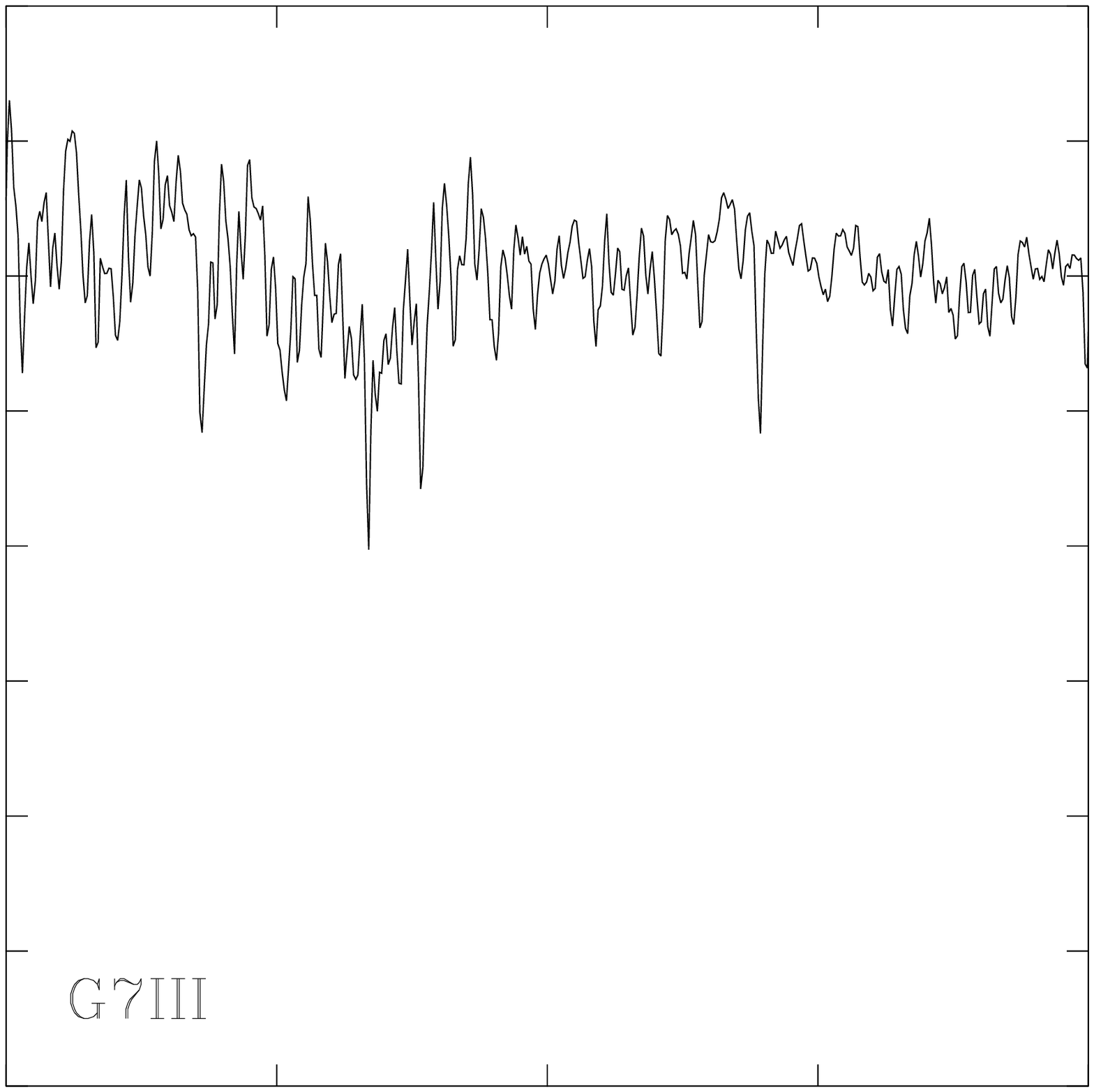}
\includegraphics[width=\figcolsize\textwidth]{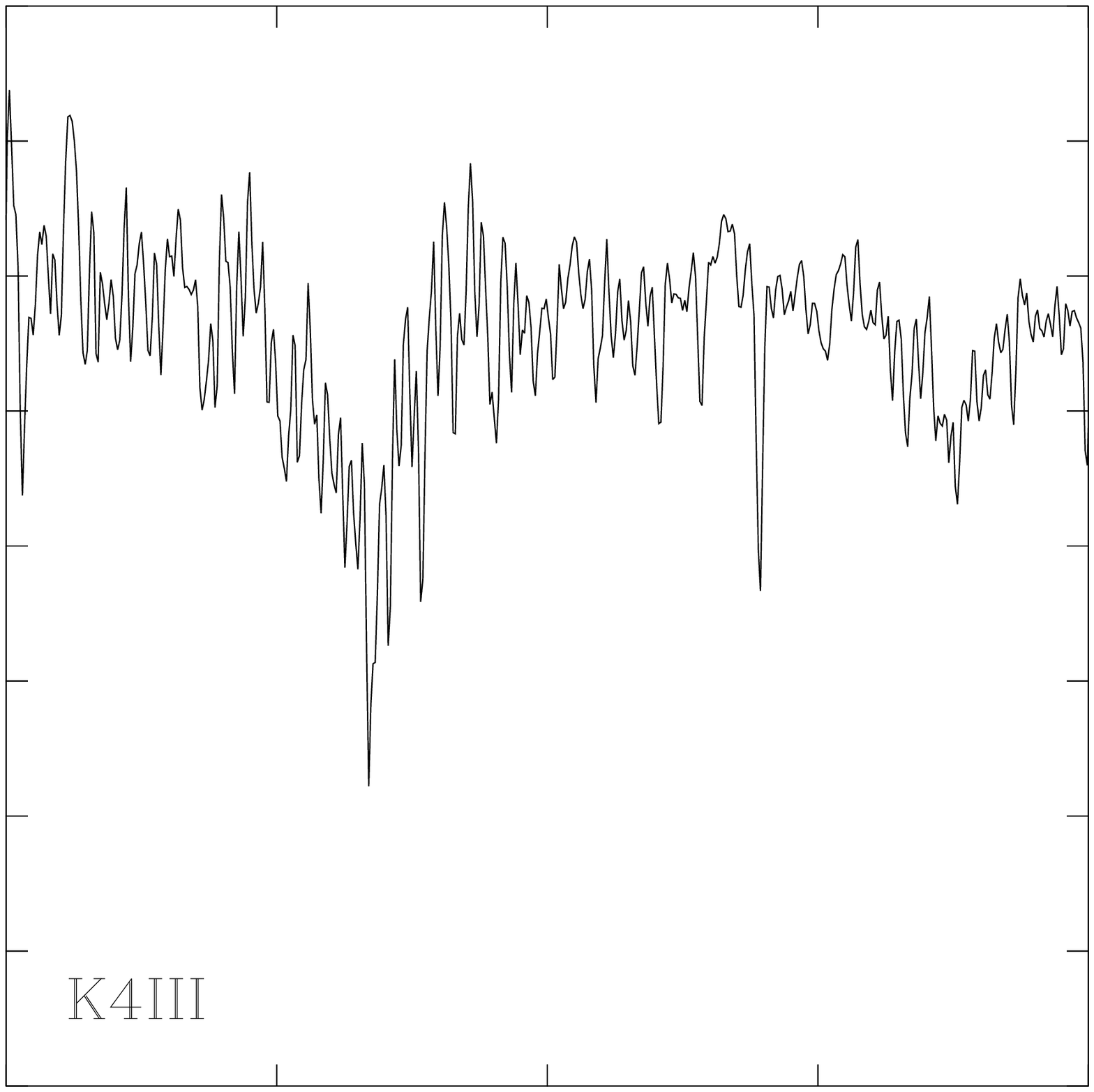}
\includegraphics[width=\figcolsize\textwidth]{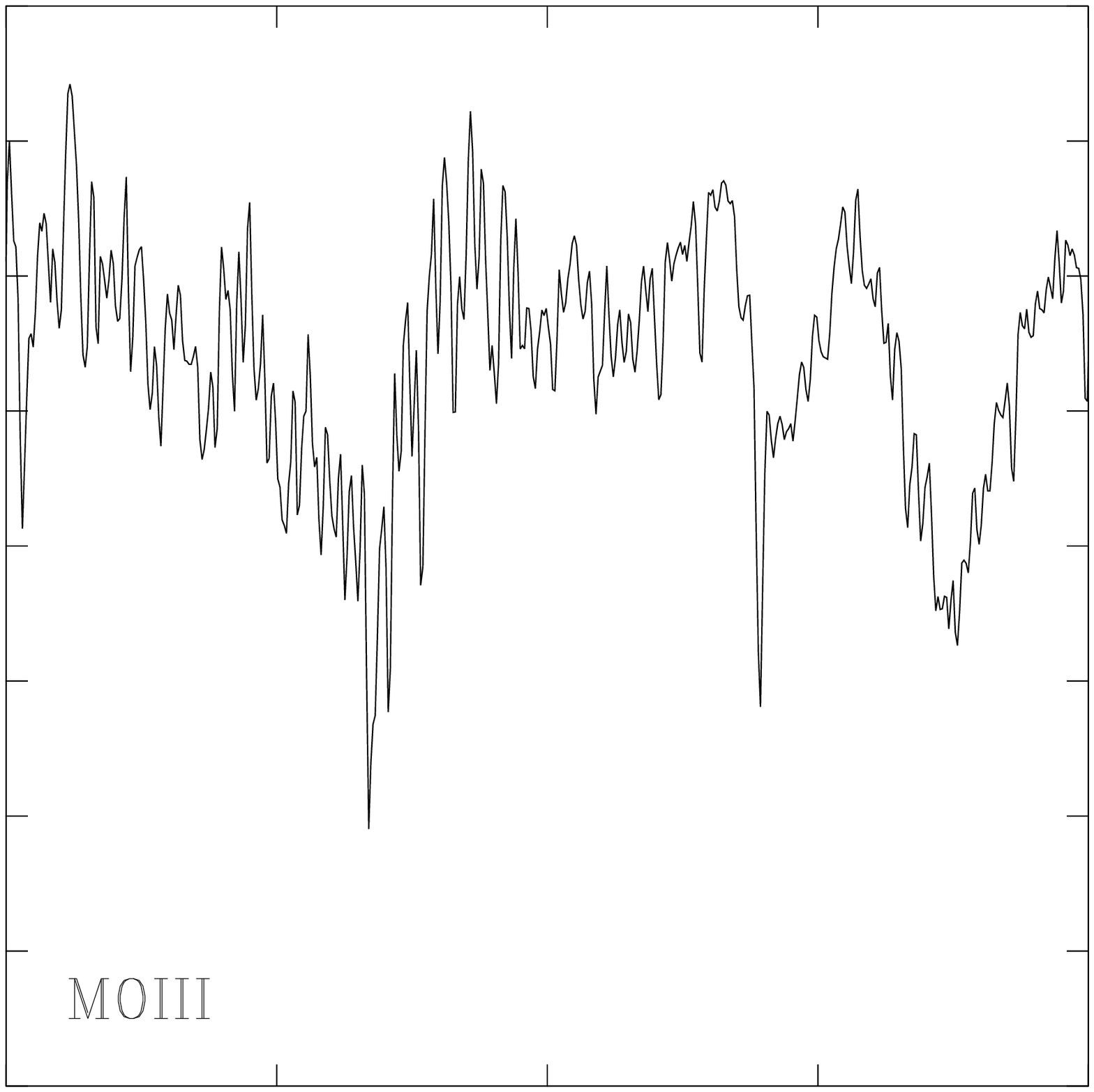}
\includegraphics[width=\figcolsize\textwidth]{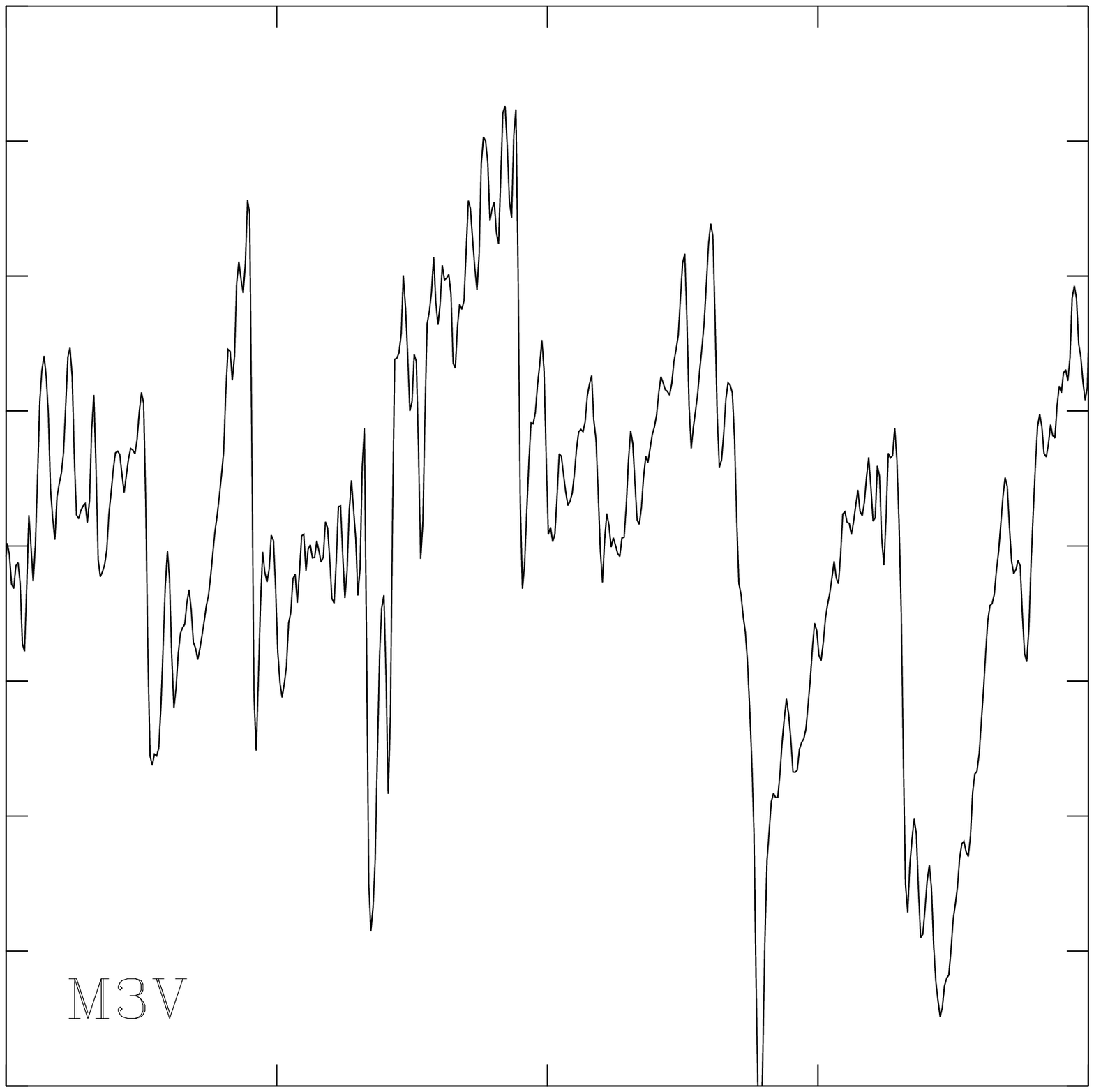}
}
\caption{Stellar spectra: sequence from early to late types derived
from \citet{valdes04}. The data have been normalised using a quadratic
fit and are plotted from 450 to 650\,nm at nm resolution;
relative intensity scale is 0.3 to 1.1}
\label{fig:spectral-types}
\end{figure*}

\begin{figure*}
\centerline{
\includegraphics[width=\figcolsize\textwidth]{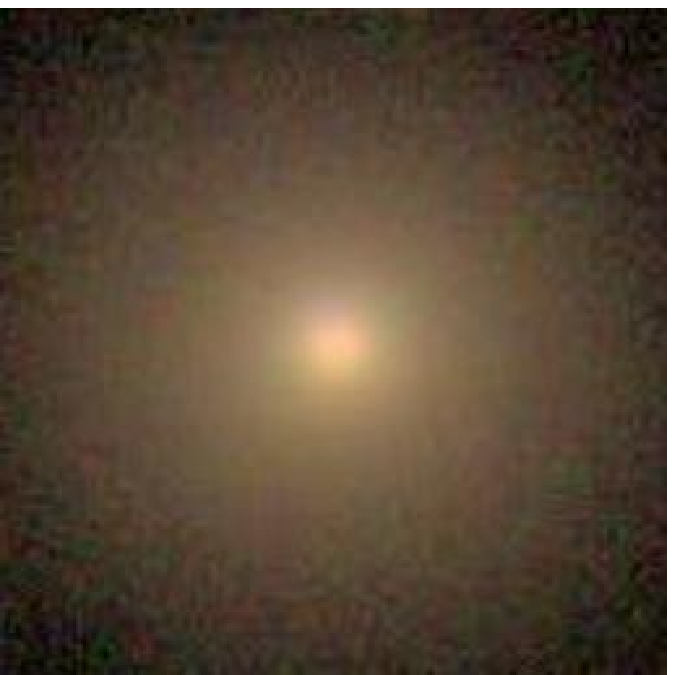}
\includegraphics[width=\figcolsize\textwidth]{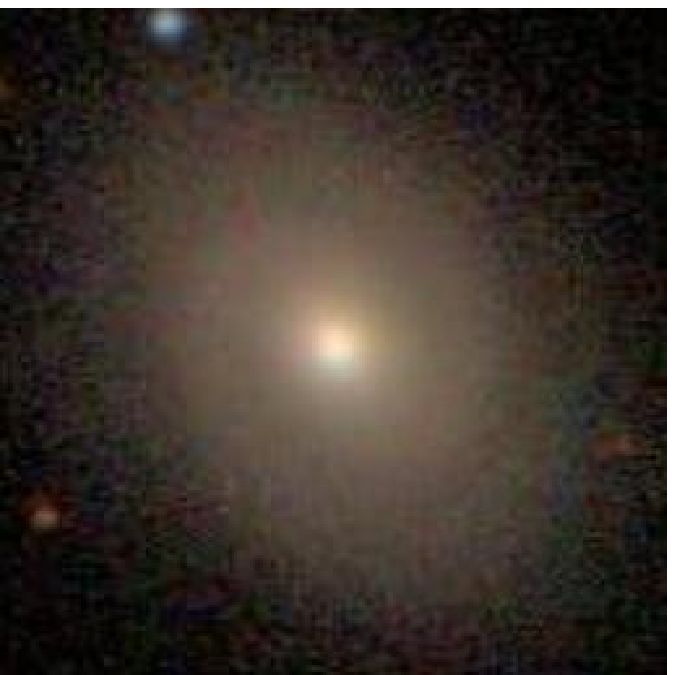}
\includegraphics[width=\figcolsize\textwidth]{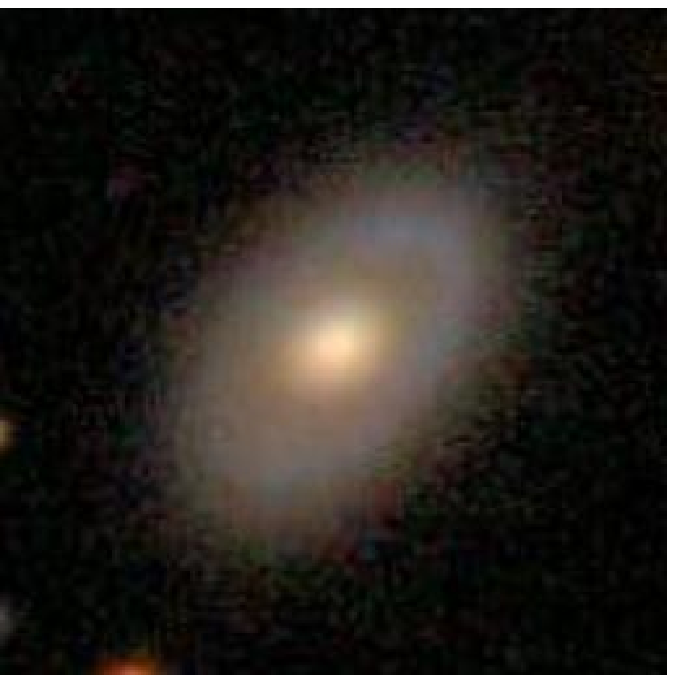}
\includegraphics[width=\figcolsize\textwidth]{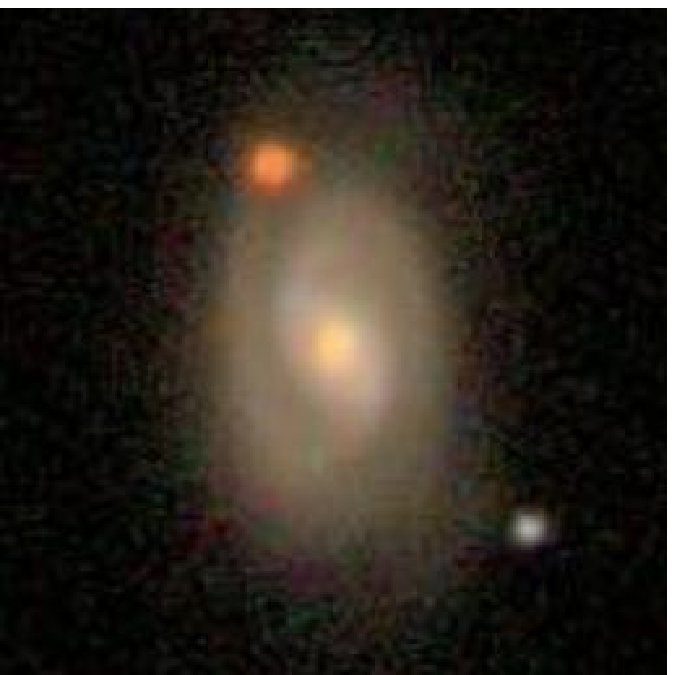}
}
\vspace{0.4cm}
\centerline{
\includegraphics[width=\figcolsize\textwidth]{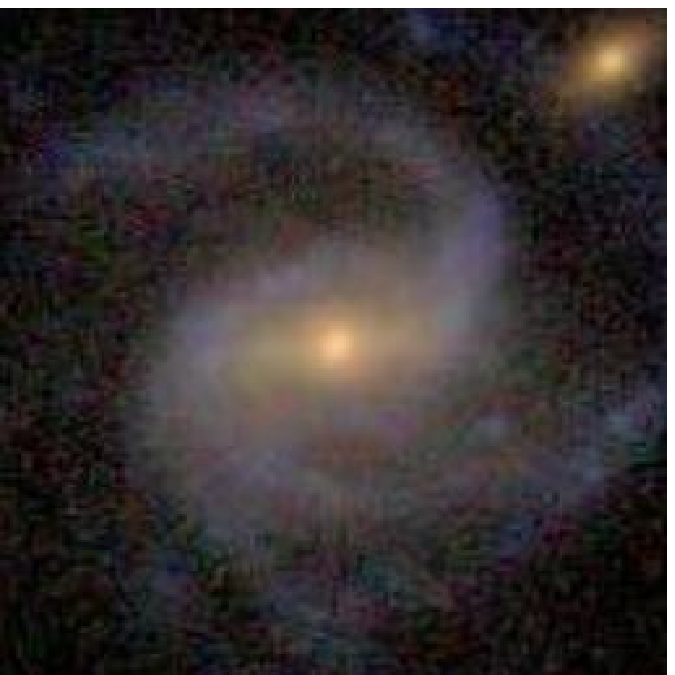}
\includegraphics[width=\figcolsize\textwidth]{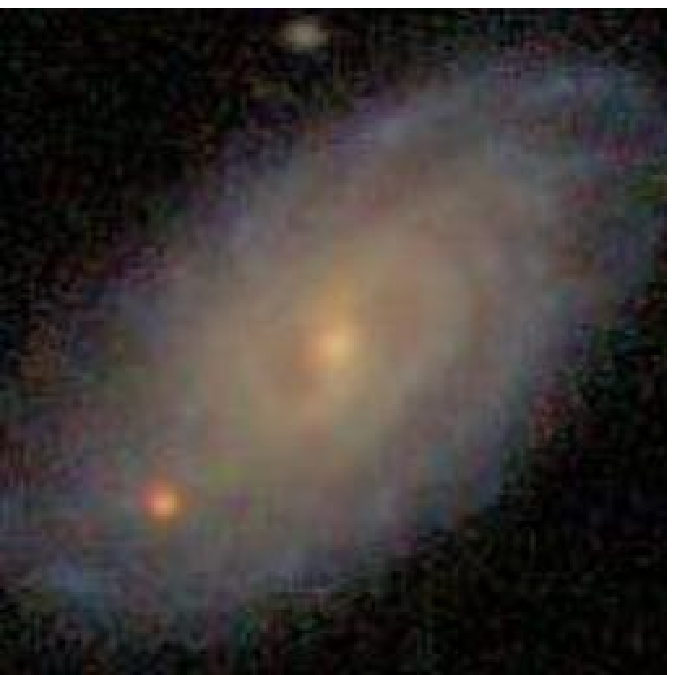}
\includegraphics[width=\figcolsize\textwidth]{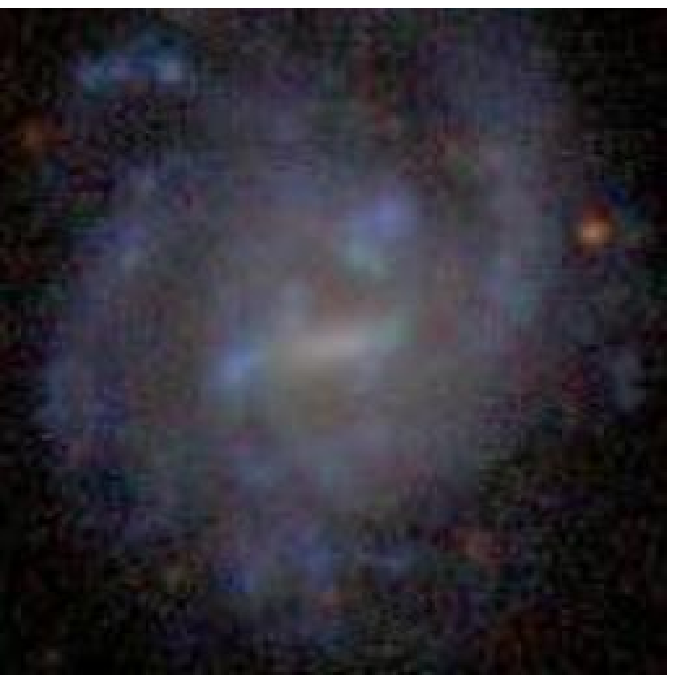}
\includegraphics[width=\figcolsize\textwidth]{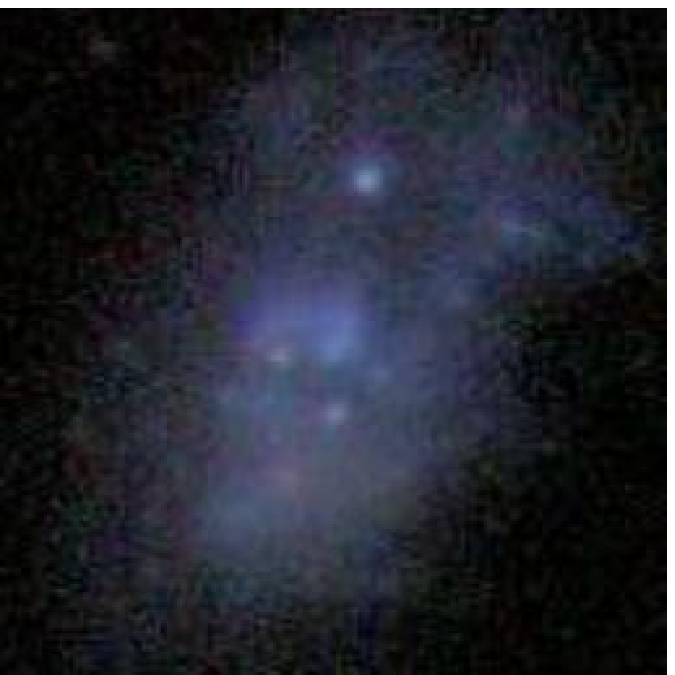}
}
\caption{Galaxy morphologies: sequence from early to late types
derived from \citet{nakamura03} classifications and Sloan Digital Sky
Survey colour images \citep{NSG04}
scaled as per \citet{lupton04}. Types are E, S0,
Sa, Sb (top row), Sc, Sc, Sd, Im (bottom row).}
\label{fig:morphology}
\end{figure*}

By the early 1920's the temporal connotations of `early' and `late' had
been largely disregarded for stellar spectra. 
Hubble knew this and used the terminology
choosing the direction of the morphological sequence based on the apparent
complexity as per stellar spectra. Figure~\ref{fig:spectral-types} shows
spectra of eight stars put in order of their spectral classification, and
Figure~\ref{fig:morphology} shows colour images of eight galaxies put in order
of their morphological classification.  From these figures, we can immediately
see the unification of the terminology such that, in general, earlier types
are simpler in appearance and later types are more complex in appearance for
both stellar spectra and galaxy images.  (It is worth bearing in mind
that both these phenomena would have been observed on black-and-white
photographic plates in the 1920's, and not intensity versus wavelength for
spectra, or composite multi-band images.)  While this definition is subjective,
it is a coherent starting point for explaining these terms in astrophysics
courses; and it is related to physical phenomena, for example, dependence of
absorption transitions on atmospheric temperature in stars, and star-formation
triggering in galaxies.

The history of science is simplified in science textbooks because their
primary aim is ``persuasive and pedagogic'' \citep{Kuhn96}. Yet the particular
inaccuracy discussed in this paper, while minor, does a disservice to Hubble
and observational astronomy, and provides no clarification. By the early 20th
century, astronomy was a mature science, and in the mid-1920's the concept of
`extragalactic' had only recently been largely accepted. Hubble's 1926 paper
is the first to use this term in a published title. Therefore, for Hubble to
preempt a theory of `galaxy evolution' by suggesting that galaxies evolved
along the sequence is an historian's fallacy.  Hubble was a careful
observational astronomer and it is quite clear from his 1926 and 1927 papers,
quoted above, that he assumed the temporal implications of `early' and `late'
had been dropped prior to his usage of them. I postulate that he would not
have presumed to establish a theory of galaxy evolution at this stage.  Even
in his more comprehensive book published a decade later \citep{Hubble36}, he
was strictly neutral with regard to evolution. He, however,
was influenced by Jeans' development of liquid rotating spheroids 
and did earlier hint at evolution based on Jeans' dynamics, even as he insisted 
that his classification was strictly based on morphology with
no interjections about origins \citep{Sandage05}.

Why not have used the terms `simple' and `complex'? These would have preempted
theory.  In fact, many morphologically classified early-type galaxies have
been shown to have complicated internal dynamics such as kinematically
decoupled cores \citep{deZeeuw02}.  The terms `simple in appearance' and
`complex in appearance' are clunky in comparison with `early' and `late', and
it should be noted that the complexity of appearance is a guide to the order
of the sequence not the definition. In the case of stellar spectra, the order
is generally quantified by the strengths of various absorption bands. For
morphological classification of galaxies, there is no consensus on the most
useful quantification. \citet{Hubble36} considered the sequence as a
``progression in dispersion or expansion'' of spiral arms. There are many
alternative strategies for galaxy classification \citep{Sandage05} but
Hubble's scheme remains part of commonly used systems today.

In summary, when introducing the terms `early' and `late' for the
morphological classification of galaxies, the historical context is explained
incorrectly in many texts.  I have shown that the logical reason relates to
the complexity of appearance within the sequence. This reason should improve a
student's grasp of why these, apparently arbitrary, terms are used for both
stars and galaxies. Rather than abandoning the terms, I propose that Hubble's
intention be kept in mind when using them since the temporal connotations
should by now be well and truly dispelled.

\section*{Acknowledgements} 

Many thanks to Allan Sandage for suggestions and corrections. 
Thanks to Meghmala Choudhury and Robin McLean for checking references,
Phil James for
suggesting I write this paper, and Nancy Medley, Sue Percival and John Stott
for comments.  The SAO/NASA Astrophysics Data System was used for historical
research, and SDSS visual tools was used for extracting the galaxy images.

%%\bibliographystyle{mn2e}
%%\bibliography{galaxies,stars,surveys}

\bsp

\label{lastpage}

\end{document}